# Hybrid Superconducting Neutron Detectors

V. Merlo[1], M. Salvato[1,2], M. Cirillo[1,2], M. Lucci[1], I. Ottaviani[1], A. Scherillo[3],

G. Celentano[4] and A. Pietropaolo[4,5*]

*1* Dipartimento di Fisica, Università Tor Vergata, Via della Ricerca Scientifica, I-00133 Roma, Italy
*2* **CNR**-SPIN, Italy
*3* Science and Technology Facility Council, ISIS Facility Chilton Didcot Oxfordshire, UK
*4* ENEA Frascati Research Centre, V. E. Fermi 45, 00044 Frascati, Italy
*5* Mediterranean Institute of Fundamental Physics

A new neutron detection concept is presented that is based on superconductive niobium (Nb) strips coated by a boron (B) layer. The working principle of the detector relies on the nuclear reaction $^{10}B + n \rightarrow \alpha + ^{7}Li$, with $\alpha$ and *Li* ions generating a hot spot on the current-biased Nb strip which in turn induces a superconducting-normal state transition. The latter is recognized as a voltage signal which is the evidence of the incident neutron. The above described detection principle has been experimentally assessed and verified by irradiating the samples with a pulsed neutron beam at the ISIS spallation neutron source (UK). It is found that the boron coated superconducting strips, kept at a temperature $T = 8$ K and current-biased below the critical current $I_c$, are driven into the normal state upon thermal neutron irradiation. As a result of the transition, voltage pulses in excess of 40 mV are measured while the bias current can be properly modulated to bring the strip back to the superconducting state, thus resetting the detector. Measurements on the counting rate of the device are presented and the future perspectives leading to neutron detectors with unprecedented spatial resolutions and efficiency are highlighted.

*Corresponding author: antonino.pietropaolo@enea.it

Neutron imaging performed with spatial resolution below 10 μm could open new and interesting perspectives for researches in medicine and biology [1,2], material research [3], forensics [4] and chemistry [5]. Conventional neutron imaging systems can achieve a spatial resolution down to 10 μm only by reducing their efficiency at percent level. Superconductors instead are well suited to conceive and design detection devices featuring spatial resolution below the above limit with efficiency well above 1%.

The capability of superconducting thin films to detect energetic charged particles has been recognized since the late 1940s, when voltage pulses were first observed in NbN strips irradiated by α particles from a Polonium source [6]. Further studies on Sn and In strips [7,8] confirmed the bolometric nature of the process. More recently, Wedenig *et al* [9] have observed voltage pulses in NbN strips irradiated with α particles from a $^{241}$Am source ($E_\alpha$= 5.5 MeV). The detection scheme, involving the creation of a hot-spot in superconducting strips, has also been explored in the context of x-ray and γ-ray detection, as discussed in Refs. [10,11]; a long narrow strip of width $w$ (see Fig. 1) is biased with a current $I$ at a value slightly below the critical value $I_c$ and is locally driven into the normal state by the energy released by the ionizing particles. In fact, the normal-state hot spot squeezes the current carrying channel where the critical current is locally exceeded causing a superconducting to normal state transition across the whole width of the strip. Evidence of the hot spot is found in a fast voltage drop of the transport current which, relying on a bolometric model, can be calculated as $V = 2r_c \rho J$, $r_c$ being the radius of the hot spot [7,8], $\rho$ the electrical resistivity of the material and $J$ the measured transport current density.

Pure metals feature small voltage pulses, because of the relatively low values of $\rho$ and $J_c$; Wedenig *et al*. [9] observed voltage pulse rise-times of the order of 250 ns and amplitudes of 10 mV in NbN, thus confirming the working principle of the detector. The idea herein presented is to use a boron converting layer to coat a superconducting strip. In the boron layer the reactions

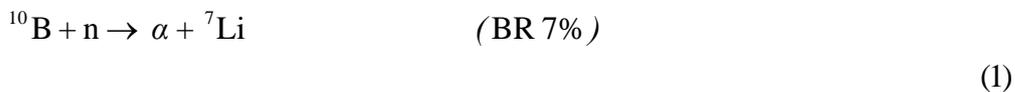

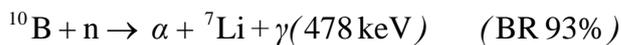

(1)

take place, where BR is the branching ratio for the particular reaction channel. Lithium and alpha particles induce hot spots in the Nb strip, while the effect of the 480 KeV γ-rays is negligible (the absorption probability is of the order of $10^{-5}$ for the Nb thickness discussed in the following). It is worth noting that, because of the kinematics of the reaction, both particles (α and Li ion) may deposit energy in the superconducting strip thus allowing for a higher detection efficiency.

The Nb strips were obtained by lift-off, depositing Nb on Si/SiO$_2$ substrates previously covered by photoresist and patterned by photolithographic techniques. It should be noted that α and Li deposit their energy also in the substrate, where phonon excitation may take place. However, heating effects via phonon propagation back to the superconducting film are deemed to be negligible because of the 1 μm thick amorphous $SiO_2$ buffer layer on top of the Si substrate, the former acting as an acoustic decoupler. This problem has been systematically tackled in Ref. [8], where it has been shown that laying a thin varnish layer on top of the substrate prior to film deposition dramatically increased the performance of the detector.

An optical image of the strip geometry, with a sketch of the four contacts and wiring, is shown in Fig. 2. For all the samples prepared, the strip length was 1 mm and the Nb thickness was 150 nm while different distances ($L$) between the voltage electrodes and width ($w$) of the strips were realized in order to test the best conditions for detection. On top of the Nb strip 450 nm of natural Boron, equivalent in terms of neutron absorption to 90 nm of $^{10}B$ - given the isotopic abundance of $^{10}B$ in natural Boron - was deposited by means of e-beam evaporation through a mechanical shadow mask. The four leads (two for the bias current and two for voltage measurements as shown in Fig. 2) were glued on the Nb contact pads by commercially available silver paint. The superconducting properties of the samples were tested by critical temperature ($T_c$) and critical current density ($J_c$) measurements obtaining typical values of $T_c$ in the range 8.5 K-9.2 K, $J_c \cong 10^{10}$ A·m$^{-2}$ at $T = 4.2$ K, and $\rho\,(T = 10$ K$) = 6.4\ 10^{-8}$ Ω·m.

All the measurements under neutron irradiation, performed on the INES beam-line [12] at the ISIS pulsed neutron source (Rutherford-Appleton Laboratory, UK) [13], were made in a $^4$He flow cryostat with the sample mounted at the tip end of a stick which also hosted a thermometer and a heater allowing to perform measurements in the range 1.8-10 K with a nominal uncertainty in temperature of 0.01 K. The neutron beam was directed along a direction normal to the sample surface and its flux was changed using remotely-controlled boron carbide ($B_4C$) jaws (see Fig. 1).

During neutron irradiation, the $I$-$V$ characteristic was monitored at a fixed temperature to get eveidence of the hot spot formation. Measurements o voltage $V$ across the strip at fixed I were also performed as a function of time. Heating effects due to the bias current through the metallic silver contacts are reduced biasing the sample at low current.

Since $I_c$ decreases with increasing temperature and vanishes at $T = T_c$, it was necessary to stabilize the sample temperature very close to $T_c$ before performing $I$-$V$ and $V$-$t$ measurements. Typical values of temperature used are $T = 0.9\ T_c$. Figure 3 shows the $I$-$V$ characteristics recorded with the neutron beam jaws closed and open for a sample with $w = 20$ μm and $L = 800$ μm. For this sample (henceforth named *sample 1*), $T_c = 8.64$ K and the $I$-$V$ measurements were performed at $T = 8.52$ K, with $T/T_c = 0.99$. As shown in Fig. 3, $I_c = 2.5$ mA and $I_c = 2$ mA with the jaws closed and open, respectively. In the latter case, a decrease of the critical current is observed of about 20%, caused by neutron absorption and subsequent charged particle release; creation of the hot-spot reduces the area available to the transport current so that, at a fixed temperature, the critical current under irradiation is lower.

In order to better support these data, Fig. 4 shows the *V-t* measurement for *sample 1* at the same temperature. The sample was biased with a current $I = 1.8$ mA, corresponding to $I/I_c = 0.72$, during the time in which the jaws are open. Before opening the jaws, the voltage is at its lowest level determined by instrumental noise, estimated to be of the order of few microvolts. The voltage remains at this level even after jaws opening, then switches at its highest level ($V = 40$ mV, corresponding to the normal state of the Nb strip) after a time $t = 60$ s. According to a procedure outlined in Ref. [7], the radius $r_c$ of the hot spot can be estimated from the difference in the critical current measured with and without irradiation obtaining a value $r_c = 8$ μm. By inserting the measured values $\rho$ and $J$, we find $V = 10$ mV, which is of the same order of magnitude as the measured voltage drop.

The same measurements reported in Figs. 3 and 4, using the device layout sketched in Fig. 2, were performed with a "bare" Nb strip, *i.e.* not coated by boron. In this case no increase of the voltage was detected for at least 30 minutes of irradiation. This negative evidence confirms the role played by the B layer on the Nb strip, ruling out the possibility of spurious effects due to the neutron interaction with Nb or with the Ag contact pads. Indeed, the contact pads could in principle alter the measurements since Ag becomes activated under neutron irradiation thus generating *β*-decay products. From the experimental evidence, it can be inferred that such spurious effects are not relevant for the measurement and that the boron layer is essential for generating the transitions of the superconducting strip.

Figure 5 shows the continued operation of a detector (*sample 2*) with $w = 10$ μm, $L = 800$ μm and 450 nm of Boron as absorber. Data were taken at $T = 8.12$ K, i.e. $T/T_c = 0.92$. The red line refers to the open-closed jaws status, which follows the on-off bias current, while the black line represents the voltage change after the superconducting-normal state transition produced by neutron detection. Upon neutron detection the bias current is driven to zero and the $B_4C$ jaws closed. After about 10 s to allow for temperature stabilization, the current is raised to its previous value and the jaws are re-opened to record a new event. As a consequence, it is possible to estimate the detector count rate. The detection events take place at different times after each jaws opening thereby confirming the statistical nature of the phenomenon.

Further analysis has been performed on *V-t* traces recorded on *sample 1*; the temperature was kept constant at $T = 8$ K and data were taken as a function of the bias current (*e.g.* one trace for each value of the current). In order to collect more events and increase statistical significance, data have been recorded for a much longer time (typically 30 minutes or more). From the *V-t* series a count rate has been measured, taking into account dead time effects (the detector is not always "ready" since it needs to be reset after each transition to the normal state). Results are shown in Fig. 6, where the count rate is plotted as a function of the bias current of the superconducting film: the slope of the weighted least-squares line is 4.47 ± 0.39 counts/(s·A).

This linear behavior has been already observed and extensively discussed in Ref. [8] where it has been shown, with proper modeling, that it can be ascribed to details of the critical current distribution within the strip; we note in passing that this affords a convenient way of choosing the operating range of the detector. We also observe a non-zero intercept upon extrapolation of our data to zero bias current; actually, the line does not extrapolate to zero, but to a value $I_0$ beyond which the count rate goes rapidly to zero; the value $I_0$ (the "onset" current) defines the lower bound of the region where particle detection becomes possible; this behavior has been observed in Ref.8. Because of the low counting statistics, our measurements were confined to a region much closer to the critical current where only the linear behavior is visible.

In order to compare our results with theoretical predictions, the detector neutron count rate can be written as follows:

$$C(t) = \int_{E\min}^{E\max} \Phi(E_0) \cdot A \cdot P_n(E_0) \cdot \overline{P}_{\alpha Li}(E_0) \cdot dE_0 \qquad (2)$$

where $\Phi(E_0)$ is the neutron spectral fluence rate onto the strip, $A$ the Boron detection area, $P_n(E_0)$ the neutron absorption probability in $^{10}$B, given by $1 - \exp[-N\sigma_a(E_0)]$, $N$ the $^{10}$B atomic density per unit area and $\sigma_a(E_0)$ the 10B neutron absorption cross section. $\overline{P}_{\alpha,Li}(E_0)$ is the α/Li ion escape probability from the Boron layer. In the energy region where the neutron flux at the INES beam line is maximum (thermal region), the energy dependence of $\overline{P}_{\alpha,Li}(E_0)$ can be neglected given the large *Q*-value of the *n*-$^{10}$*B* reactions as compared to neutron energy triggering the reaction (1). This allows to simplify the relation (still holding with a high degree of accuracy in the thin neutron absorption layer approximation) that reads:

$$C(t) = NA\overline{P}_{\alpha,Li} \int_{E\min}^{E\max} \Phi(E_0) \cdot \sigma(E_0) dE_0 \quad . \qquad (3)$$

Following Ref. [14], where results for the same geometry herein considered are presented, $\overline{P}_{\alpha,Li} \approx 0.5$.

We are now in condition to compare the maximum measured count rate in our experiment , of the order of 3 counts/min, By insertimg into Eq. (3) the values of the neutron flux provided by INES beam line neutron beam monitor and $^{10}$B neutron absorption cross section values taken from standard databases [15], $C(t) = 9$ count/min is estimated, in theassumption that either Li or α are always detected. Taking into account the uncertainty on the neutron flux data and geometrical factors, the comparison with our experimental data that provide $C(t) = 3$ count/min is acceptable. The efficiency at the thermal peak has been thus estimated to be 0.5%.

Finally, it is worth stressing that there is significant room for improvement in this detection scheme towards a finer spatial resolution: the strip width can be reduced down to 1 μm using optical techniques-based lithography. In principle, strips of sub-micrometer width could be patterned, but in this case one should make use of electron beam lithography or complex optical instrumentation. Also, as far as the spatial resolution is concerned, it would be interesting to generate matrices capable of providing position in the plane and this is a challenging issue to be addressed. It has to be alsoremarked that the detection efficiency at fixed width, presently roughly less than 1%, can be increased using thicker $^{10}$B layers. Moreover, the use of superconductors with transition temperatures higher than that of Nb (9.2 K) ( such as NbN) could improve the ease in cryogenic handling of the detectors.

**Acknowledgements**

The authors thank the ISIS sample environment group for support in the cryogenics and Gareth Howells from the Computing Group. Support from E.M. Schooneveld and N.J. Rhodes (ISIS detector group) is also acknowledged for making equipment available in their laboratory.

M.C. acknowledges the financial support of CNR-SPIN

V.M and M.S. also acknowledge financial support within the CNR-STFC agreement.

**Figure Captions**

**Figure 1**: Sketch of the working principle of our hybrid neutron detection system: a superconductive niobium strip is driven normal by the hot-spot originated by alpha particles or lithium nuclei which are in turn produced by a nuclear reaction in the boron coating film caused by the incident neutron beam.

**Figure 2**: Optical image of the Nb strip with the four contact pads; also drawn are the silver paint drops and the wires of the four-probes measurements.

**Figure 3**: *I-V* characteristics at $T$ = 8.52 K measured with the jaws closed (black line) or open (red line) for the $w$ = 20 μm, $L$ = 800 μm strip detector (*Sample 1*). The critical currents are $I_c$ = 2.5 mA and 2 mA, respectively.

**Figure 4**: Voltage measurement (scale on the left) as a function of time with the jaws opened at $t$ = 0 for the *Sample 1*. The temperature was $T$ = 8.52 K and the current bias (scale on the right) was $I$ = 1.8 mA; the current rise and the jaws opening happen at the same time. The increase of the voltage at $t$ = 60 s is due to the Nb transition from the superconducting to the normal state due to neutron interaction with the device.

**Figure 5**: *V-t* measurement for the $w$ = 10 μm, $L$ = 800 μm strip detector with 450 nm of natural Boron, equivalent in terms of neutron absorption to 90 nm of $^{10}$B, in the continuous operation mode. The red line (scale on the right) illustrates the on-off bias current which switches simultaneously to the open-close jaws status. The black line (scale on the left) is the measured voltage strip across the strip.

**Figure 6**: The measured count rate as a function of the bias current through the superconducting strip; the straight line is a weighted least-squares fit to the data.

Figures

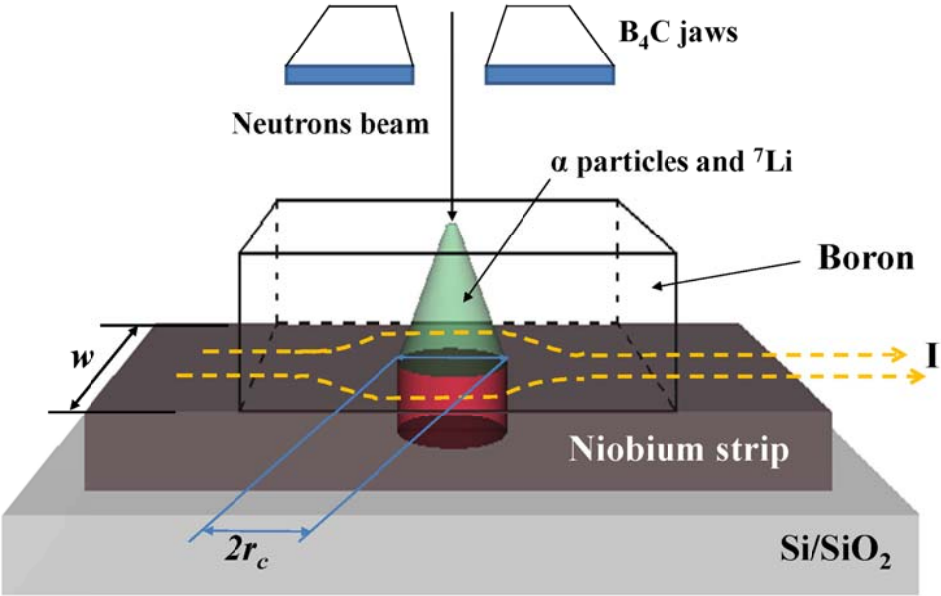

Fig. 1

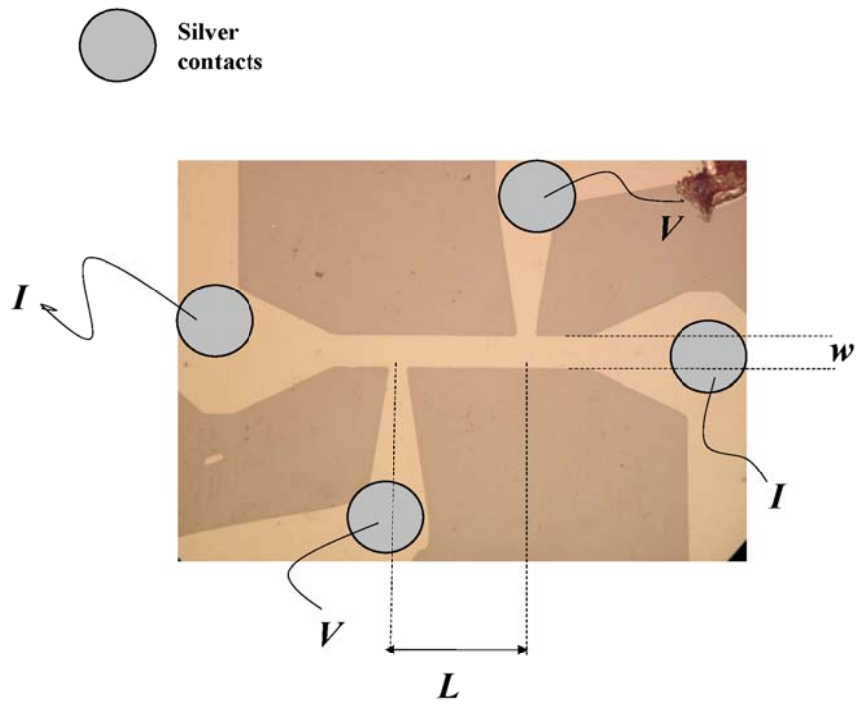

Fig. 2

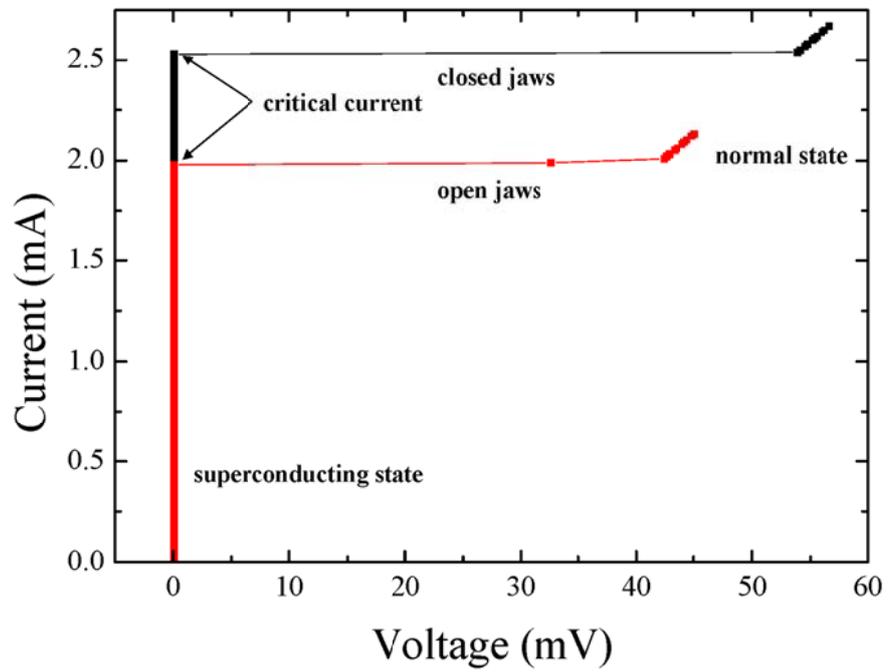

Figure 3

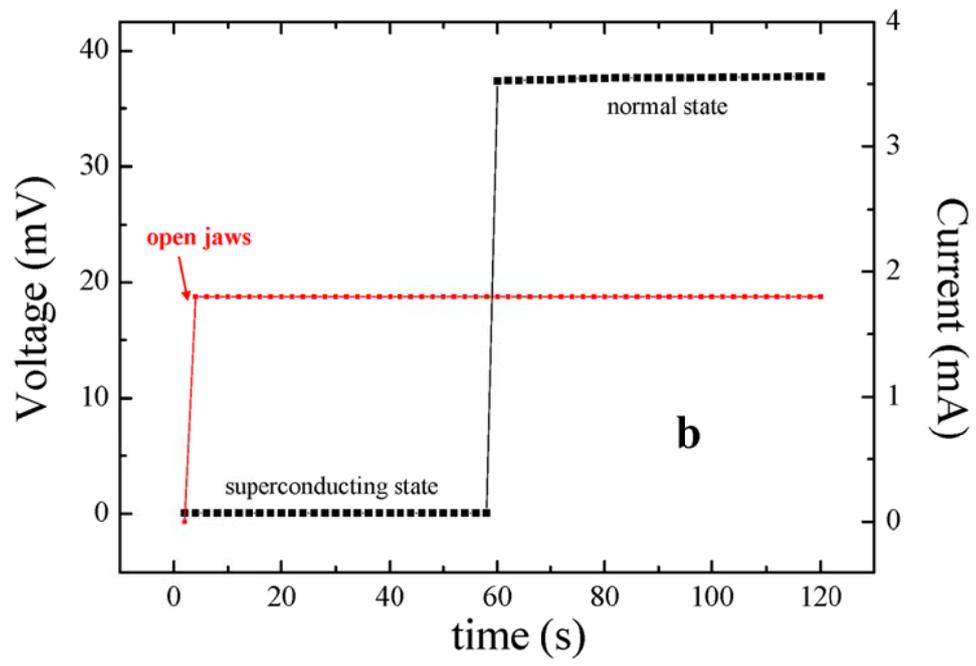

Figure 4

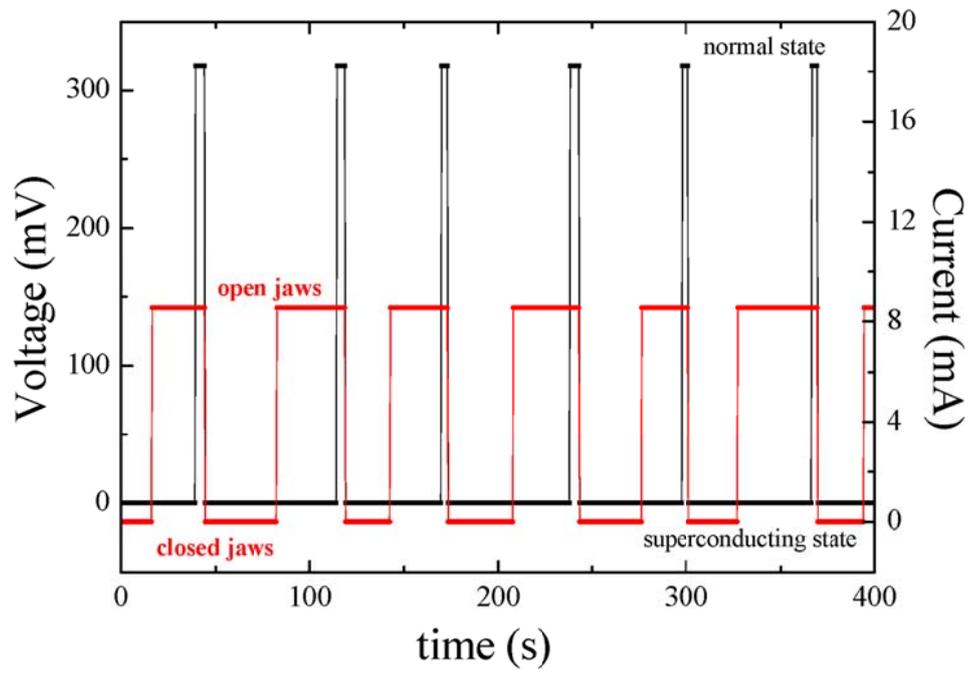

Figure 5

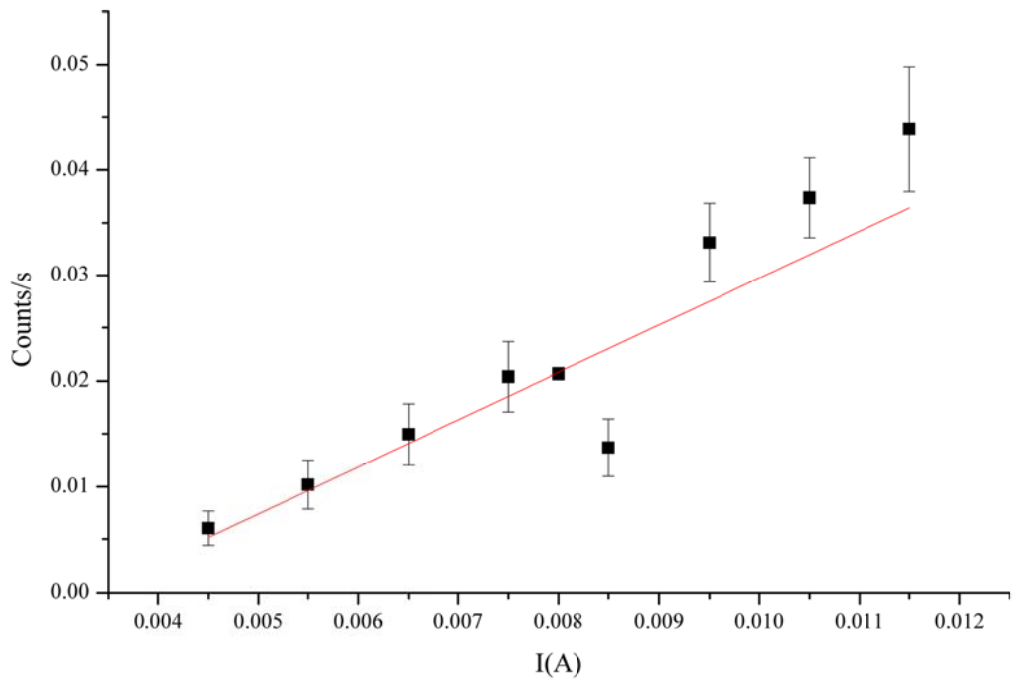

Fig. 6